\documentclass[aip,apl,twocolumn,reprint,preprintnumbers,amsmath,amssymb]{revtex4-1}

\usepackage{color}
\usepackage{graphicx}
\usepackage{dcolumn}
\usepackage{bm}

\begin{document}

\title{Search for spin gapless semiconductors: The case of inverse Heusler
compounds}

\author{S. Skaftouros}
\affiliation{Department of Materials Science, School of Natural
Sciences, University of Patras,  GR-26504 Patra, Greece}

\author{K. \"{O}zdo\u{g}an}
\affiliation{Department of Physics, Yildiz Technical University,
34210 \.{I}stanbul, Turkey}

\author{E. \c{S}a\c{s}{\i}o\u{g}lu}
\affiliation{Peter Gr\"{u}nberg Institut and Institute for
Advanced Simulation, Forschungszentrum J\"{u}lich and JARA, 52425
J\"{u}lich, Germany} \affiliation{Department of Physics, Fatih
University, 34500, B\"{u}y\"{u}k\c{c}ekmece, \.{I}stanbul, Turkey}

\author{I. Galanakis}\email{galanakis@upatras.gr}
\affiliation{Department of Materials Science, School of Natural
Sciences, University of Patras,  GR-26504 Patra, Greece}

\begin{abstract}
We employ \emph{ab-initio} electronic structure calculations to
search for spin gapless semiconductors, a recently identified new
class of materials, among the inverse Heusler compounds. The
occurrence of this property is not accompanied by a general rule
and results are materials specific. The six compounds identified
show semiconducting behavior concerning the spin-down band
structure and in the spin-up band structure the valence and
conduction bands touch each other leading to 100\% spin-polarized
carriers. Moreover these six compounds should exhibit also high
Curie temperatures and thus are suitable for spintronics
applications.
\end{abstract}


 \maketitle

The rapid growth of nanotechnology the last two decades brought to
the attention of scientific research novel materials with novel
properties. Especially the design of new magnetic nanomaterials
leaded the development of spintronics.\cite{ReviewSpin} To achieve
the incorporation of spin in electronic devices there are two
ways: either the development of magnetic materials at the
nanoscale, e.g. the growth of half-metallic ferromagnetic Heusler
compounds, \cite{ReviewHM} or the doping of semiconductors with
magnetic ions like in diluted magnetic
semiconductors.\cite{ReviewDMS} All these developments were
triggered by the development  of computational materials science,
which made accessible the simulation of compounds in order to
predict the electronic, magnetic, optical,... properties. Moreover
simulations allow to study known alloys in new metastable
structures where their properties are completely altered with
respect to the known stable lattice structures.

\begin{figure}
\includegraphics[width=\columnwidth]{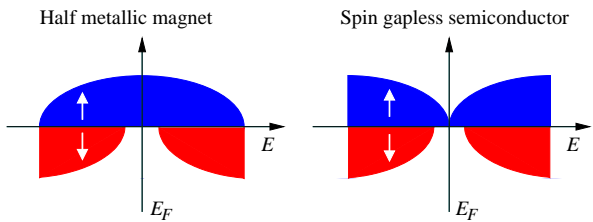}
\caption{(Color online) Schematic representation of the density of
states of a half-metallic system (left panel) and of a spin
gapless semiconductor (right panel). With blue (red) we show the
spin-up (spin-down) states.} \label{fig1}
\end{figure}

A special class of materials incorporated the so-called gapless
semiconductors; semiconductors with vanishing gap width.\cite{GS}
These materials are of special interest since the mobility of
carriers is considerably larger than usual semiconductors. The
first gapless semiconductors studied were Hg-based IV-VI
compounds, such as HgCdTe, HgCdSe and HgZnSe, but these alloys are
toxic and easily oxidized.\cite{GS} Afterwards, PbPdO$_2$ was
proposed to be a gapless semiconductor,\cite{Kurzman11} and the
zero gap-width was also demonstrated experimentally.\cite{Chen11}
The most known gapless semiconductor is the
graphene.\cite{graphene} Wang in 2008 proposed that the doping of
PbPdO$_2$ would lead to a new class of materials which he named
spin gapless semiconductors (SGS).\cite{Wang} SGS is the border
between half-metallic (HM) ferromagnets\cite{deGroot} and
semiconductors and a schematic representation of the density of
states (DOS) is shown in Fig.\,\ref{fig1}. In a half-metallic
ferromagnet the spin-up (majority-spin) band is crossed by the
Fermi level as in a usual magnetic metal while in the spin-down
(minority-spin) band a gap exists and the Fermi level falls within
the gap as in semiconductors. In the case of SGS the picture of
the spin-down band is similar to the HM but now in the spin-up
band the Fermi level falls within a zero-width gap. The system can
be still magnetic since the two spin-band structures are
different. The advantage of SGS is that no-energy is required to
excite electrons from the valence to the conduction band, and
excited carriers (both electrons and holes) can be 100\%
spin-polarized simultaneously leading to new functionalities of
spintronics devices. Other SGS materials include (i) the graphene
ribbons altered by CH$_2$ radical groups,\cite{Pan11} where the
magnetism stems from the unsaturated carbon states, (ii) the
ferromagnetic semiconductor HgCr$_2$Se$_4$ which becomes SGS under
a pressure of 9 GPa,\cite{Liu12} and (iii) the BN nanoribbons with
vacancies.\cite{Pan10}

Heusler compounds is a huge family of compounds with more than
1000 members exhibiting a variety of diverse magnetic
phenomena.\cite{FelserRev} Almost all compounds crystallize in a
cubic close-packed structure similar to the zinc-blende structure
of binary semiconductors. Recently, new phenomena have been
identified in some of these Heusler groups like topological
insulators,\cite{Chadov10} half-metallic
ferromagnets,\cite{deGroot,ReviewJPD} half-metallic
antiferromagnets,\cite{HMA,Galanakis11,Hu} multifunctional
Heuslers for recording,\cite{Kurt} etc. Thus one could easily
imagine the possibility of finding also Heusler compounds being
SGS among them. It seems that this is true among the so-called
inverse full-Heusler compounds of the X$_2$YZ chemical formula
crystallizing in the so-called XA structure with space group
$F\overline{4}3m$; the prototype being CuHg$_2$Ti. The lattice is
actually a fcc one with four atoms as basis along the diagonal
occupied in the sequence X-X-Y-Z. X and Y are transition metal
atoms and for the XA structure to occur the  valence of X should
be lower than the valence of Y. Examples are
Cr$_2$MnSb,\cite{Galanakis09} Mn$_2$CoGa,\cite{Mn2CoGa}
Cr$_2$CoGa,\cite{Galanakis11} etc. Recently, Ouardi and
collaborators identified Mn$_2$CoAl to be a SGS and experimental
evidence was given towards this direction.\cite{Ouardi12} Heusler
compounds compared to other potential SGSs have the advantage of
being compatible with current semiconductors technology, their
magnetic properties are stable and there is a continuously growing
knowhow on the synthesis of such alloys in the form of
films.\cite{FelserRev} In this Letter we have used the
full-potential nonorthogonal local-orbital minimum-basis band
structure scheme (FPLO)\cite{koepernik} within the  generalized
gradient approximation (GGA)\cite{gga} to study the electronic and
magnetic properties of all the inverse X$_2$YZ alloys where X=Sc,
Ti, V, Cr or Mn, Y= Al, Si or As and Y is a transition-metal atom
ranging from Ti to Zn. First we determined the equilibrium lattice
constants using total energy calculations and a dense
20$\times$20$\times$20 $\mathbf{k}$-point grid to carry out the
numerical integrations. Our results have shown that within this
large number of Heuslers studied there are six of them being SGS:
Ti$_2$MnAl, Ti$_2$CoSi, Ti$_2$VAs, V$_3$Al, Cr$_2$ZnSi and
Mn$_2$CoAl. The present Letter focuses on these six alloys which
can be incorporated as SGS in spintronics devices.

As we just mentioned we have identified six inverse full-Heusler
compounds which are SGSs and in Table\,\ref{table1}, we have
gathered the calculated equilibrium lattice constants and magnetic
moments. First, to calculate the equilibrium lattice constants we
have performed total energy calculations scanning a wide range of
lattice constant values and taking a more dense grid around the
one minimizing the total energy. Then we have fitted a quadratic
polynomial using the values close to the energy minimum in order
to better estimate it. The use of GGA is crucial since the
local-spin-density-approximation (LSDA) is well-known to
underestimate the equilibrium lattice constants with respect to
experimental values. The calculated values, as shown in
Table\,\ref{table1} exceed the 6 \AA\ for the Ti-based alloys and
V$_3$Al while they are considerably smaller for Cr$_2$ZnSi and
Mn$_2$CoAl. This is expected since Ti and V atoms have large
atomic radius and their \emph{d}-wavefunctions extend more in
space with respect to Cr and Mn atoms. Between the two latter
compounds Cr$_2$ZnSi has the larger equilibrium lattice constant
due to the presence of the larger Zn atom with respect to the Co
one in Mn$_2$CoAl. The experiments by Ouardi \emph{et al.} on
Mn$_2$CoAl gave a lattice constant of 5.798 \AA ,\cite{Ouardi12}
which deviates less than 1.2\% from our estimated values of 5.73
\AA.

\begin{table}
\caption{Calculated  equilibrium lattice constant and spin
magnetic moments in $\mu_B$ for the inverse Heusler compounds
under study. We use the symbols A and B to denote the two early
transition metal atoms sitting at different sites (see text for
explanation). Note that the total spin magnetic moment is given
per formula unit (which coincides with the per unit cell value).
The last column is the sum of the absolute values of the
atom-resolved spin magnetic moments.}
\begin{ruledtabular}
\begin{tabular}{llcccccc}
X$_2$YZ & a(\AA ) & $m^{X(A)}$ & $m^{X(B)}$ & $m^Y$ & $m^Z$ &
$m^{total}$ & $m^{abs}$ \\ \hline
Ti$_2$MnAl  & 6.24 & 1.44  & 1.30 & -2.74 & -0.01 & 0 & 5.49\\
Ti$_2$CoSi  & 6.03 & 1.80  & 0.86 &  0.35 & -0.02 & 3 & 3.03\\
Ti$_2$VAs   & 6.23 & 1.31  & 0.53 & -1.86 &  0.01 & 0 & 3.71\\
V$_3$Al     & 6.09 & 1.64  & 0.00 & -1.64 &  0.00 & 0 & 3.28\\
Cr$_2$ZnSi  & 5.85 & -1.89 & 1.93 &  0.01 & -0.05 & 0 & 3.88\\
Mn$_2$CoAl  & 5.73 & -1.65 & 2.80 &  0.94 & -0.08 & 2 & 5.47
\end{tabular}
\end{ruledtabular}
\label{table1}
\end{table}

Also in Table \ref{table1}, we present the atom-resolved spin
magnetic moments in $\mu_B$ and the total one per formula unit
which for these compounds coincides with the per unit cell value.
We use the superscripts A and B to distinguish the two X atoms
sitting at the A and B sites (for the structure see
Ref.\,\onlinecite{Galanakis09}). We should first comment on the
total spin magnetic moments. For all compounds the obtained values
are integers. This is expected since due to the gaps the total
number of occupied states in both spin directions is integer and
thus the difference between the spin-up and spin-down occupied
states, which equals to the total spin magnetic moment in $\mu_B$,
is also an integer. The four alloys Ti$_2$MnAl, Ti$_2$VAs, V$_3$Al
and Cr$_2$ZnSi  exhibit a zero total spin magnetic moment and thus
can be also classified as half-metallic antiferromagnets also
known as half-metallic fully-compensated
ferrimagnets.\cite{Galanakis11} The other two compounds,
Ti$_2$CoSi and Mn$_2$CoAl, show a net total spin magnetic moment
of 3 and 2 $\mu_B$ respectively. This behavior of the total spin
magnetic moment follows a generalized version of the
Slater-Pauling behavior of the Heusler
compounds\cite{Galanakis02a,Galanakis02b}  and will be discussed
elsewhere.\cite{GSP} The advantage of the half-metallic
antiferromagnetic compounds is that they create no external fields
and thus lead to minimal energy losses.

The atom resolved spin moments do not show a coherent behavior
among all compounds.  To understand their behavior we have to take
into account the so-called Bethe-Slater curve.\cite{Bethe-Slater}
The Cr and Mn atoms when nearest neighbors tend to couple
antiferromagnetically and this explains the relative orientation
of the spin magnetic moments for Cr$_2$ZnSi and  Mn$_2$CoAl
presented in Table\,\ref{table1}. The Zn atom has all its
\emph{d}-states filled and only the delocalized 4\emph{s}
electrons contribute to magnetism leading to a negligible Zn spin
magnetic moment  in Cr$_2$ZnSi. Co couples ferromagnetically to
its nearest-neighboring Mn$^B$ atom due to the hybridization of
the \emph{d}-\emph{d} orbital as occurs also in the usual
full-Heusler compounds containing Co and Mn.\cite{Galanakis02b} In
the case of the Ti-based compounds the two Ti atoms have parallel
spin magnetic moments as expected by the Bethe-Slater curve. The
spin moments of Mn and V atoms are antiparallel to the spin
moments of the Ti atoms while the spin moment of the Co atom is
ferromagnetically coupled to the one of the Ti atoms in Ti$_2$CoSi
compound. V$_3$Al is a special case since the V atoms at the A and
C sites have antiparallel spin magnetic moments of the same
magnitude while the V atom at the B site and the As atom have no
net spin magnetic moment. The two V atoms at the A and C sites are
equivalent by symmetry,\cite{Galanakis02b} and thus are only
allowed to have either parallel or antiparallel spin magnetic
moments of the same magnitude. The second configuration is the
ground state as our total energy calculations suggest. By symmetry
the V atoms at the B site has as nearest neighbors eight V atoms,
four at the A site and four at the C site, and thus by symmetry
should have a zero spin magnetic moment. The \emph{sp} atoms at
the Z sites have very small spin magnetic moments in all cases.
The experiments of Ouardi et al in Ref. \onlinecite{Ouardi12} gave
for Mn$_2$CoAl a total spin magnetic moment of 2 $\mu_B$ as
predicted by our calculations, while their first-principles
calculations gave for the atom-resolved spin magnetic moments
about -2, 3 and 1 $\mu_B$ for the Mn$^A$, Mn$^B$ and Co atoms in
agreement with our results.

\begin{figure}
\includegraphics[width=\columnwidth]{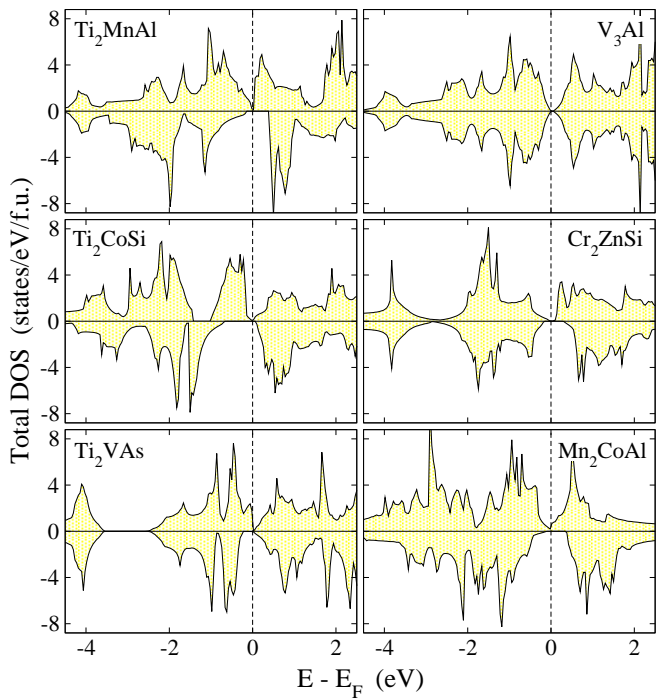}
\caption{(color online) Total density of states (DOS) per formula
unit for the compounds under study. Positive DOS values correspond
to the majority-spin (spin-up) states and negative DOS values to
the minority-spin (spin-down) states. The zero of the energy axis
corresponds to the Fermi level.} \label{fig2}
\end{figure}

\begin{figure}
\vskip -1cm
\includegraphics[scale=0.33,angle=270]{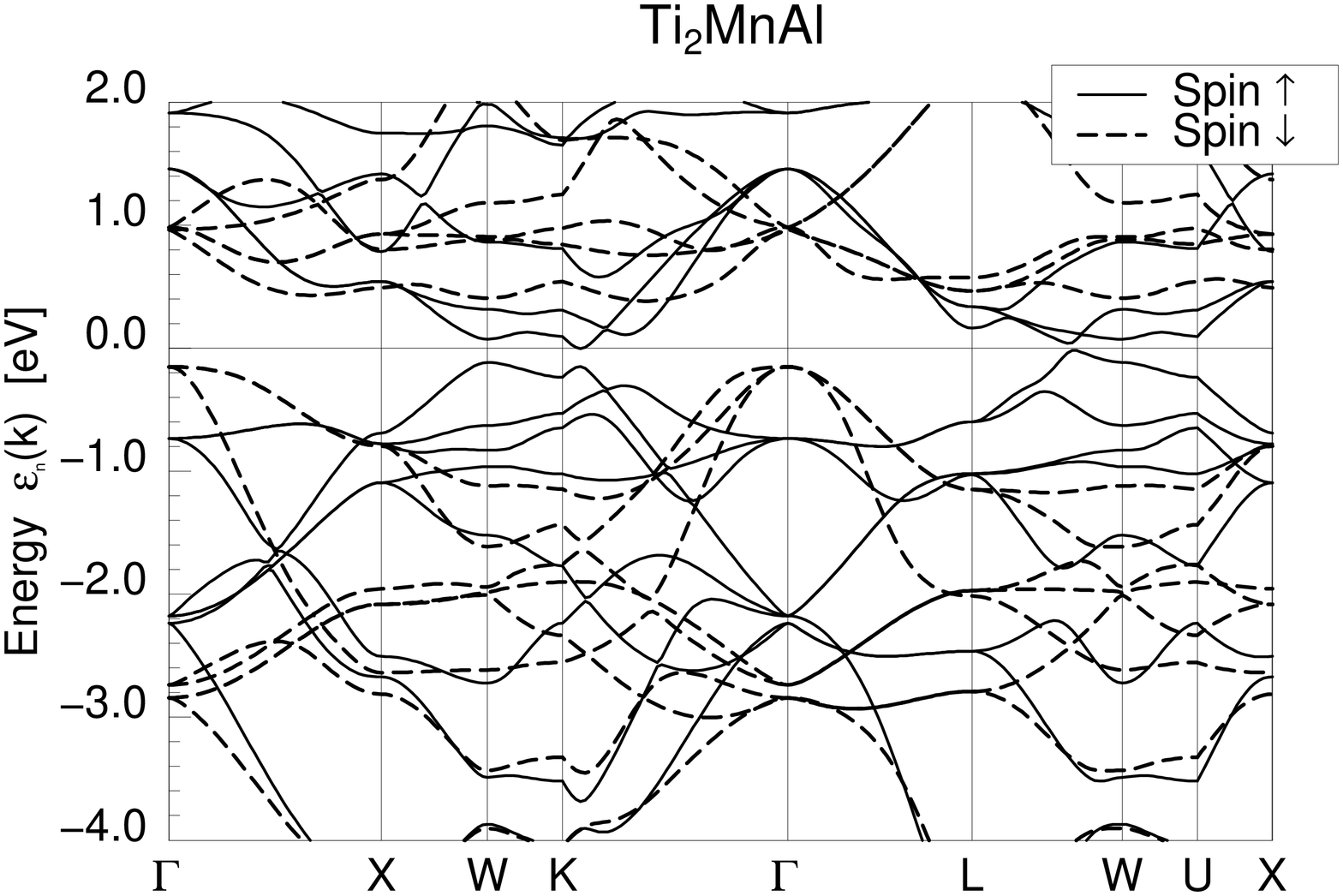} \vskip
-1cm
\includegraphics[scale=0.33,angle=270]{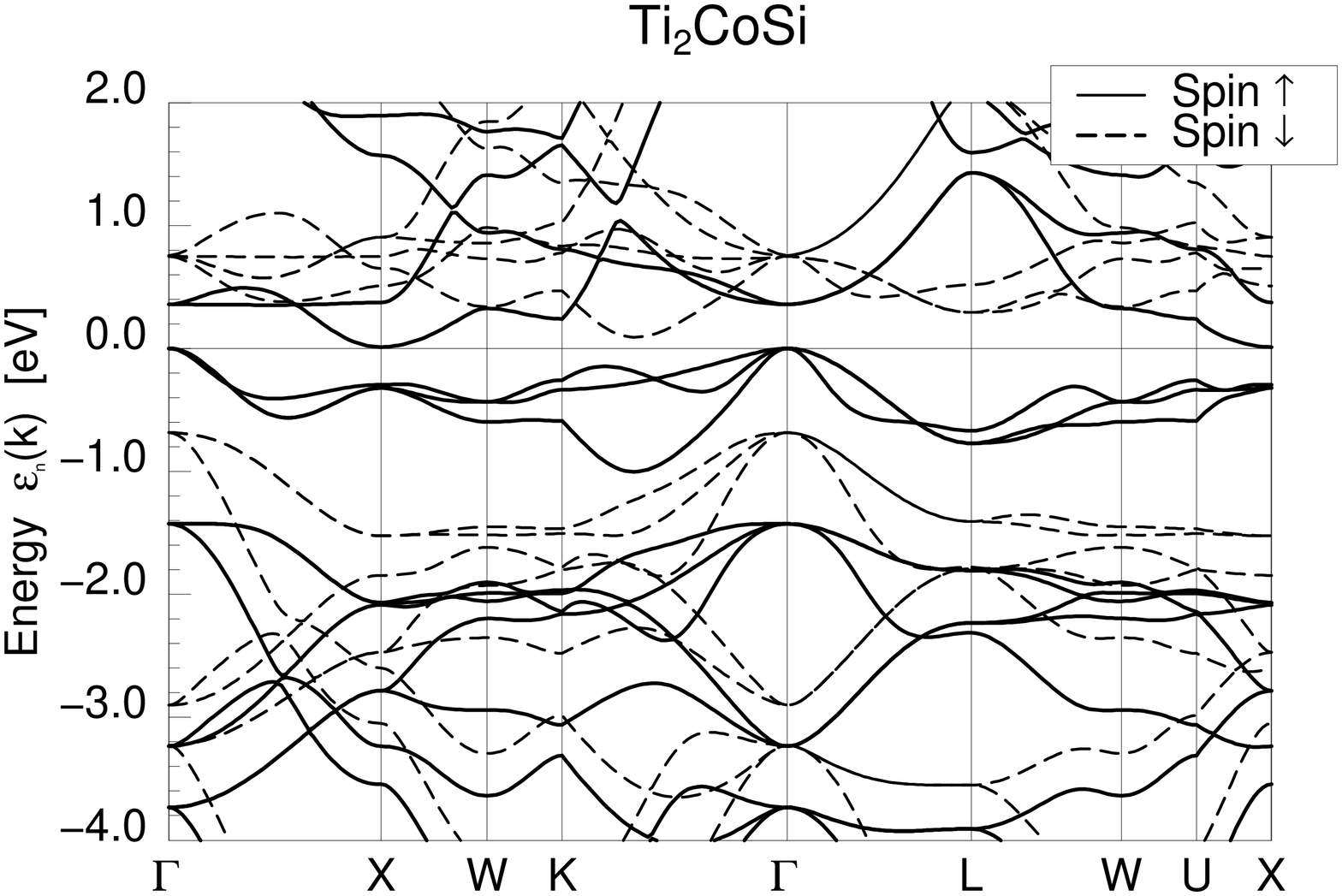}
\vskip -0.5cm \caption{ Band structure along the high-symmetry
directions for the Ti$_2$MnAl (upper panel) and Ti$_2$CoSi (lower
panel) compounds. The zero energy value in the vertical axis
corresponds to the Fermi energy. With the solid [dashed] lines we
present the majority-spin (spin-up) [minority-spin (spin-down)]
electronic bands.} \label{fig3}
\end{figure}

Now let us discuss the main finding of our study: the appearance
of spin gapless semiconducting behavior in these six compounds. In
Fig.\,\ref{fig2} we present the total density of states (DOS) for
all six compounds. Positive DOS values concern the majority-spin
or spin-up states and negative DOS values the minority-spin or
spin-down states (in the case of zero total spin magnetic moments
the terms majority-spin and minority spin are meaningless). In all
six cases there is a sizeable gap in the spin down band structure
and the Fermi level falls within this gap. In the spin-up band
structure the valence and conduction bands touch each other and
the Fermi level falls within a zero-width gap. Thus these
compounds can be classified as SGSs. The width of the spin-down
gap is larger for Ti$_2$CoSi which has the highest total spin
magnetic moment of 3 $\mu_B$, since in that case the exchange
splitting between the two spin bands is expected to be strong.
Although the total spin magnetic moment for Mn$_2$CoAl is also
important, the antiferromagnetic coupling of the neighboring
spin-magnetic moments leads to a spin-down gap comparable to the
other half-metallic antiferromagnetic compounds. V$_3$Al is a
special case since the spin-down and spin-up bands are identical.
If we examine the states close to the spin-up gap we can see that
they do not have the same behavior in all cases, \textit{e.g.} in
Ti$_2$MnAl they are more steep than in Cr$_2$ZnSi.

The question which arises is if there is a common feature in the
band structure of these compounds which could lead to an on-demand
design of these materials. We have plotted the spin-dependent band
structure for all six compounds. The Mn$_2$CoAl band structure is
similar to the one calculated in Ref.\,\onlinecite{Ouardi12} with
a $\Gamma$-X indirect gap in the spin-up band. In Fig.\,\ref{fig3}
we show the band-structure for Ti$_2$MnAl and Ti$_2$CoSi focusing
around the Fermi level. In the case of Ti$_2$MnAl we have a direct
gap in the midpoint between the L and W high-symmetry points. In
the case of Ti$_2$CoSi we have an indirect gap with the maximum of
the valence spin-up states at the $\Gamma$ point and the minimum
of the conduction spin-up states at the X-point like for
Mn$_2$CoAl. Overall the band structures are quite different and no
safe conclusion can be drawn. Also in the case of Mn$_2$CoAl if we
replace Ga or In for Al, which belong to the same column of the
periodical table, the Mn$_2$CoGa and Mn$_2$CoIn alloys are no more
SGS and are simple half-metallic ferrimagnetis.\cite{Liu08} It
seems that the appearance of a zero gap in the spin-up band
structure is a rare phenomenon which is material dependent and no
general rules exist concerning the design of new SGSs.

Finally, we would like to comment on the expected Curie
temperature, $T_\mathrm{C}$, of these compounds which is crucial
for applications. Previous extensive studies on multi-sublattice
half-metallic Heusler compounds have shown that the Curie
temperature is more or less proportional to the total spin
magnetic moment (or sum of the absolute values of the atomic spin
magnetic moments in the case of ferrimagnets)  since the
$T_\mathrm{C}$ is mainly determined by nearest neighbor
inter-sublattice exchange interactions.\cite{Tc1,Galanakis11,Tc2}
Among the considered compounds Mn$_2$CoAl was found experimentally
to have a Curie temperature of 720 K in
Ref.\,\onlinecite{Ouardi12} while the sum of the absolute values
of the spin moments (see last column of Table \ref{table1}) is
5.47 $\mu_B$. Based on this empirical rule we expect a
$T_\mathrm{C}$ of 400 K for Ti$_2$CoSi which has the lowest sum.
An interesting case is Ti$_2$MnAl which combines a zero total spin
moment with a very high Curie temperature as expected from the
very large sum of the absolute values of the atomic spin moments
which is almost identical to Mn$_2$CoAl.

Employing \emph{ab-initio} electronic structure calculation, we
have identified six spin gapless semiconductors among the inverse
Heusler compounds: Ti$_2$MnAl, Ti$_2$CoSi, Ti$_2$VAs, V$_3$Al,
Cr$_2$ZnSi and Mn$_2$CoAl. All six compounds show a semiconducting
behavior in the spin-down band while in the spin-up band structure
the valence and conduction bands touch each other. No general rule
can be stated to predict the occurrence of this property. The
behavior of the spin magnetic moments can be explained based on
the Bethe-Slater curve. Moreover, we expect these alloys to
exhibit Curie temperatures exceeding the room temperature and thus
they are suitable for spintronics applications since both type of
carriers, electrons and holes, are spin-polarized and can be
separately manipulated. We expect the results in this Letter to
trigger further experimental research, as in
Ref.\,\onlinecite{Ouardi12} towards the growth of such
nanostructures.

\end{document}